# Synthesis of Hg-based cuprate superconductors
# HgBa$_2$Ca$_{n-1}$Cu$_n$O$_{2(n+1)+\delta}$ by CsCl flux additional method

Hiroshi Hara[1,2*], Ryo Matsumoto[1,2], Shintaro Adachi[1], Aichi Yamashita[1,2], Hiroyuki Takeya[1], and Yoshihiko Takano[1,2]

[1]*MANA, National Institute for Materials Science (NIMS), 1-2-1 Sengen, Tsukuba, Ibaraki 305-0047, Japan*

[2]*Graduate School of Pure and Applied Sciences, University of Tsukuba, 1-1-1 Tennodai, Tsukuba, Ibaraki 305-8577, Japan*

E-mail: HARA.Hiroshi@nims.go.jp

We have succeeded in the synthesis of Re-doped HgBa$_2$Ca$_{n-1}$Cu$_n$O$_{2(n+1)+\delta}$ quickly by a sealed tube technique with a bit of CsCl flux. Powder X-ray diffraction measurement revealed that the single phases of Re-doped HgBa$_2$CuO$_{4+\delta}$ (Hg1201), HgBa$_2$CaCu$_2$O$_{6+\delta}$ (Hg1212), and HgBa$_2$Ca$_2$Cu$_3$O$_{8+\delta}$ (Hg1223) formed in 4 h. The sample grains exhibited plate-like morphology with the sizes of 20×20×1 μm$^3$ and the chemical compositions of Hg1201, Hg1212, and Hg1223. The superconducting transitions were observed at 25 K for Hg1201, 122 for Hg1212, and 133 K for Hg1223 in the magnetic susceptibility. These results indicate that the developed method is useful for the fabrication of Hg-based superconducting wires.

Mercury-based cuprate superconductor (HgSC) of $HgBa_2Ca_{n-1}Cu_nO_{2(n+1)+\delta}$ ($n$=1, 2, 3, ...), especially $HgBa_2Ca_2Cu_3O_{8+\delta}$ (Hg1223), exhibits the highest superconducting transition temperature ($T_c$) among superconductors under ambient pressure. Such high $T_c$ makes HgSC attractive for fundamental studies to elucidate a mechanism of high-$T_c$ superconductivity and also for technical applications. However, a synthesis of HgSC is more difficult than other cuprate superconductors due to the toxicity and high volatility of mercury. It is, hence, necessary to synthesize them in a closed system such as under high pressure or in a sealed tube. In a high pressure method, there is a consensus to obtain HgSC single phases but this method produces only a few milligrams of the samples and requires a large-scale apparatus.[1–4] On the other hand, a preparation procedure of a sealed tube method is simpler than that of a high pressure way, yet it is not easy to synthesize HgSC single phases because synthesis parameters are complexly involved each other. Especially, heating temperature and partial pressures of mercury and/or oxygen greatly affect growth of HgSC single phases.[5–8] Therefore, it is essential to exploit simple synthesis method for HgSC to develop fundamental studies and technical applications in HgSC.

In this study, we report the successful synthesis of Re-doped $HgBa_2CuO_{4+\delta}$ (Hg1201), $HgBa_2CaCu_2O_{6+\delta}$ (Hg1212), and Hg1223 single phases quickly by the sealed quartz tube method with a bit of a flux. Cesium chloride (CsCl) was the chosen flux which is often used for a single crystal growth in a sealed quartz tube for, e.g., iron-based superconductor of SmFeAs(O,F),[9] bismuth chalcogenide-based superconductors of RE(O,F)BiCh$_2$ (Ch = S, Se; RE = La, Ce, Pr, Nd),[10–12] and so on. We tried the single crystal growth of Hg1201, Hg1212, and Hg1223 by the CsCl flux method, but it was not succeeded. Therefore, we reversed an amount ratio of a flux and a material: not a little material into much flux but a little flux into much material. By the new idea, the single phases of Hg1201, Hg1212, and Hg1223 were synthesized in 4 h.

The synthesis of Hg1201, Hg1212, and Hg1223 was based on a sealed quartz ampoule technique. The schematic drawing of the sample preparation is given in Fig. 1. First, the precursor powders having the compositional ratios of Re:Ba:Ca:Cu = 0.2:2:$n$−1:$n$ were synthesized by a solid-state reaction method. Starting materials of $ReO_2$, $BaCO_3$, $CaCO_3$, and CuO were mixed in the above-mentioned compositions and heated at 880−920ºC for 10−20 h in air with several intermediate grindings. The calcined precursor (~0.3 g) was mixed with HgO (~0.3 g), pelletized and put into an evacuated quartz ampoule. Moreover, a pellet of CoO (~0.28 g) as an oxygen getter was placed into the same ampoule for well promoting the growth of Hg-based superconducting phases (Fig. 1(a)).[5-7] The ampoule was heated at 780−820ºC for 12 h. The obtained material was ground and mixed with HgO (~0.3 g) again with or without a bit of CsCl (~0.03 g). The mixture was charged into an $Al_2O_3$ tube, which was inserted into an evacuated quartz ampoule. A pellet of CoO (~0.28 g) was also placed into the same ampoule (Fig. 1(b)). The ampoule was heated at 660−840ºC for 4−60 h followed by furnace cooling.

Powder X-ray diffraction (XRD) pattern was measured by $\theta - 2\theta$ method with a CuK$\alpha$ radiation using Mini Flex 600 (Rigaku). The sample morphology and the chemical composition were identified by a scanning electron microscope (SEM) using JSM6010-LA (JEOL) equipped with an energy dispersive X-ray analyzer (EDX) at the accelerated voltages of 5 kV and 20 kV. The magnetic susceptibility was measured under an applied field of 10 Oe using a magnetic property measurement system (MPMS; Quantum Design) with a superconducting quantum interference device magnetometer.

Figure 2 shows the XRD patterns of Hg1223 synthesized at 840ºC with CsCl for 4 h and without CsCl for 4 h and 60 h. The single phase of Hg1223 formed in the 4 h synthesis with CsCl. The peak of CsCl was also found in the XRD pattern because CsCl was not removed from the sample. In the synthesis without CsCl, on the other hand, it was necessary for long heating time of 60 h to obtain the almost single phase of Hg1223. The short heating time of 4 h mainly produced the unreacted precursor phases of $BaCuO_2$ and Hg1212. Therefore, CsCl can shorten the reaction time and promote the growth of Hg1223 phase. Another suggestive results on the function of CsCl are shown as follows. Figure 3(a) shows the XRD pattern of Hg1212 synthesized without CsCl at various heating temperatures for 12 h. At the low heating temperature of 660ºC, Hg1201 appeared as a main phase and Hg1212 was a secondary phase. The amount of Hg1212 phase increased with increasing heating temperature up to 740ºC. At the high heating temperature of 820ºC, the Hg1212 did not form and the precursor phase of $BaCuO_2$ mainly appeared. On the other hand, the Hg1212 phase formed in 4 h at the low heating temperature of 660ºC and existed up to 740ºC using CsCl (Fig. 3(b)). Moreover, the Hg1223 formed as a major phase at 820ºC in spite of the nominal composition of Hg1212. All of the results indicate that CsCl can enhance the reaction rate of HgSC phases, which is similar to the effect reported in the previous work.[13] Furthermore, the appearance of three phases of Hg1201, Hg1212, and Hg1223 can be controlled by the nominal composition, the heating time and temperature, and CsCl addition. The XRD patterns of Hg1201, Hg1212, and Hg1223 synthesized with CsCl are summarized in Fig. 4.

Figures 5(a)-5(c) show the SEM images of Hg1201, Hg1212, and Hg1223 synthesized with CsCl. The synthesized samples displayed plate-like crystals with sizes of 20×20×1 $\mu m^3$

and impurity grains. The EDX analysis of these crystals revealed that the chemical compositions normalized by the Ba element were estimated to be $Hg_{0.7}Re_{0.25}Ba_2Cu_{1.0}O_{3.4}$ for Hg1201, $Hg_{0.87}Re_{0.25}Ba_2Ca_{1.0}Cu_{2.2}O_{6.0}$ for Hg1212, and $Hg_{0.87}Re_{0.22}Ba_2Ca_{1.9}Cu_{3.0}O_{8.3}$ for Hg1223, which are in good agreement with the ideal compositions of $Hg_{0.8}Re_{0.2}Ba_2Ca_{n-1}Cu_nO_{2(n+1)+\delta}$. Moreover, the elements of Cs and Cl were not detected in the EDX spectra of each crystal and were found in that of the impurity grains. The single crystals of Hg1201, Hg1212, and Hg1223 were grown, hence, CsCl worked as a flux intrinsically.

Figures 6 shows the temperature dependence of magnetic susceptibilities for the as-grown Hg1201, Hg1212, and Hg1223 synthesized with CsCl. All of the data exhibit clear superconducting transitions at 25 K for Hg1201, 122 K for Hg1212, and 133 K for Hg1223. The post-annealing for the obtained Hg1212 and Hg1223 at 300–400ºC overnight under oxygen atmosphere did not change the $T_c$s, suggesting that the oxygen deficiencies in $Hg(Re)O_\delta$ plane did not exist in them due to the high Re-doping.[14] The Hg1212 and Hg1223 exhibit the nearly highest $T_c$ for each phase, compared to the previous data,[15] which indicates that each phase is in nearly optimal doped state. On the other hand, the obtained Hg1201 exhibits the lower $T_c$ of 25 K than the highest one of 97 K. Since the Re-substitution introduces much oxygen into the $Hg(Re)O_\delta$ plane to stabilize the crystal structure,[16] it can make the Hg1201 the over-doped state. It is suggested that the carrier doping level of Re-substituted Hg1201 depends on the amount of Re: the higher amount of Re-substitution leads the $T_c$ of Hg1201 to be lower.[17,18] Actually, the Re-substituted Hg1201 with low amounts of Re (~0.05) synthesized using the developed method showed $T_c$ of 85 K.

In conclusion, we have successfully synthesized the single phases of Hg1201, Hg1212, and Hg1223 in 4 h by the sealed quartz tube method with a bit of CsCl flux. The XRD measurement revealed that CsCl can shorten the reaction time and promote the growth of mercury-based cuprate superconducting phases. Therefore, the developed way is useful to fabricate Hg-based superconducting wires using, e.g., a powder-in-tube method.

**Acknowledgments**

The authors thank M. Abe, M. Oishi, S. Harada, and T. Ishiyama for their support.

## Figure Captions

**Fig. 1.** Schematic drawing of the sample preparation in (a) first and (b) second sinters.

**Fig. 2.** XRD patterns of Hg1223 synthesized at 840ºC with CsCl for 4 h and without (w/o) CsCl for 4 h and 60 h.

**Fig. 3.** (a) XRD pattern of Hg1212 synthesized without CsCl at various heating temperatures (*T*) for the heating time (*t*) of 12 h. (b) XRD pattern of Hg1212 synthesized with CsCl at various *T*s for *t* = 4 h.

**Fig. 4.** XRD pattern of Hg1201, Hg1212, and Hg1223 synthesized with CsCl at *T* = 800−840ºC for *t* = 4 h.

**Fig. 5.** SEM images of (a) Hg1201, (b) Hg1212, and (c) Hg1223 synthesized with CsCl.

**Fig. 6.** Temperature dependence of magnetic susceptibility for Hg1201, Hg1212, and Hg1223 synthesized with CsCl under field cooling (FC) and zero-field cooling (ZFC) with an external field of 10 Oe.

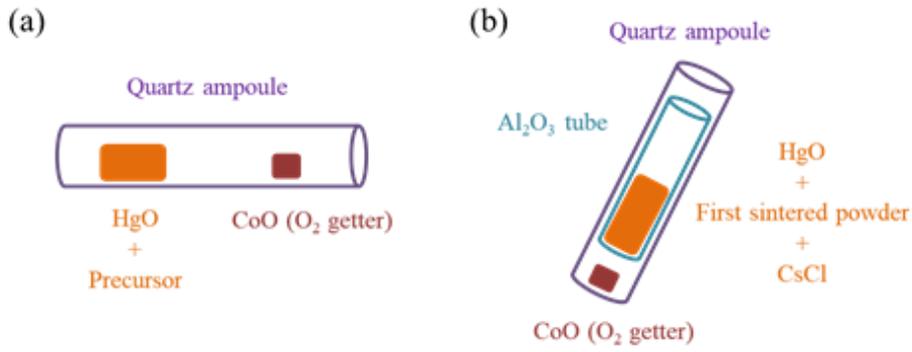

Fig. 1.

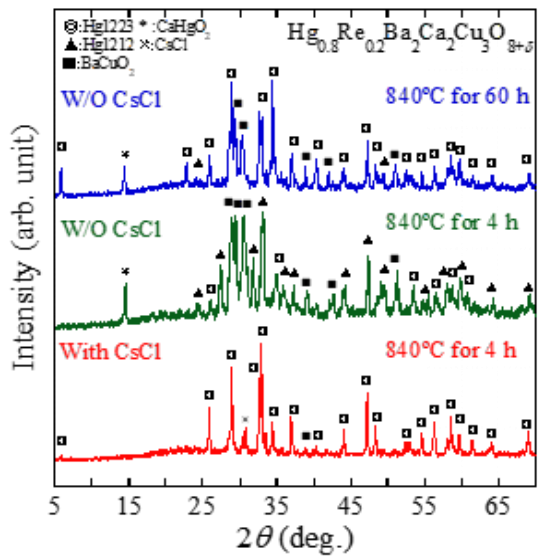

Fig. 2.

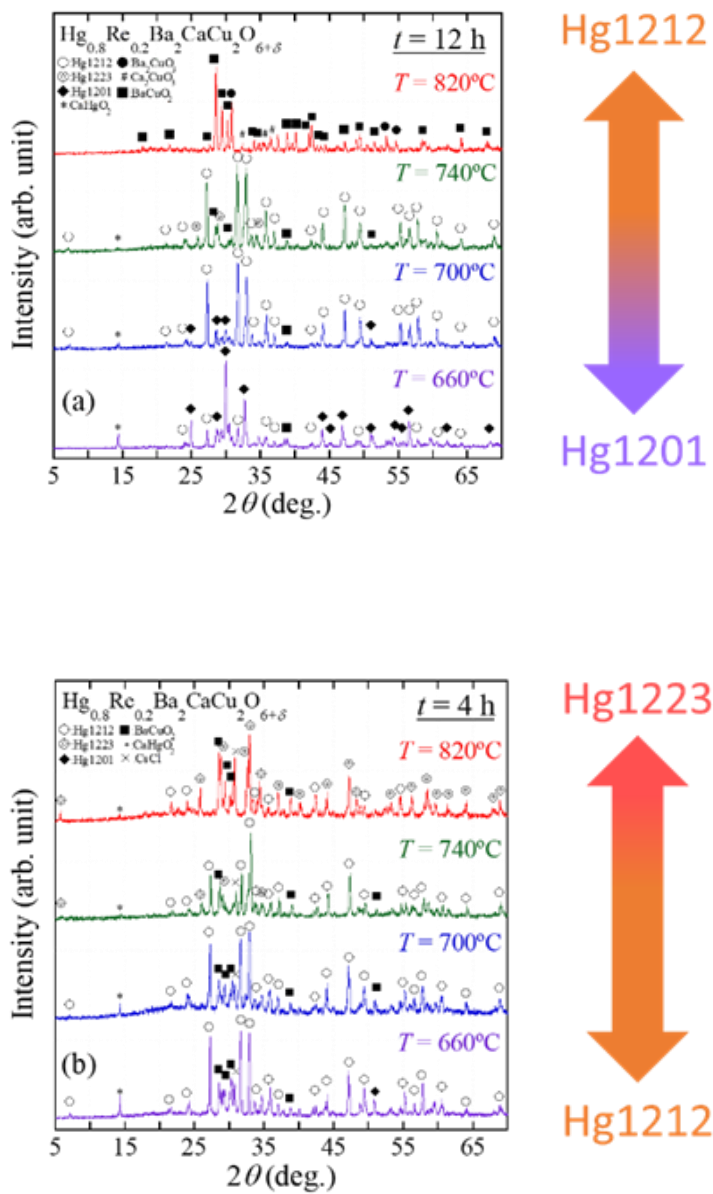

Fig. 3.

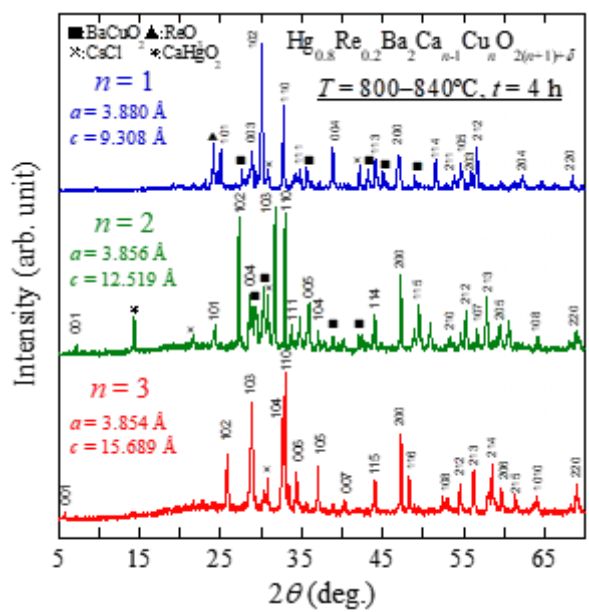

Fig. 4.

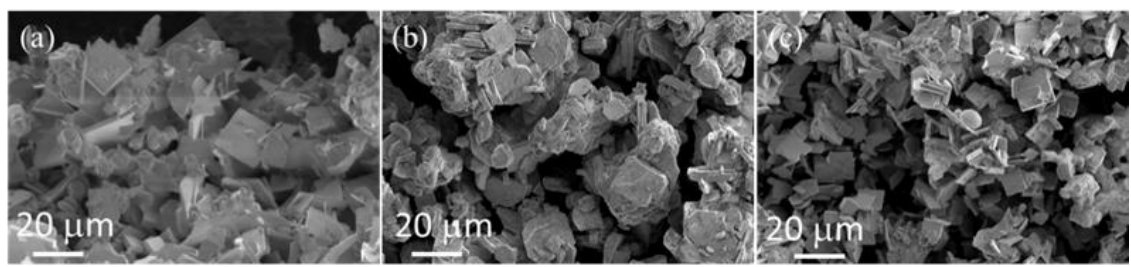

Fig. 5.

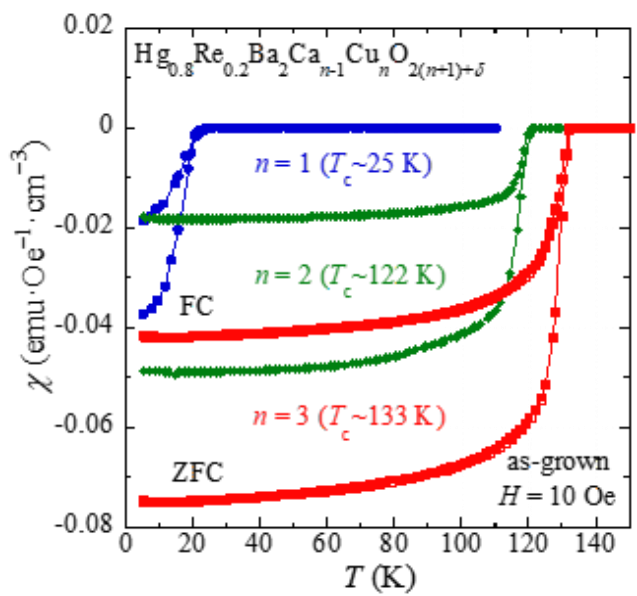

Fig. 6.